\documentclass[pdflatex,sn-mathphys-num]{sn-jnl}

\usepackage{graphicx}
\usepackage{multirow}
\usepackage{amsmath,amssymb,amsfonts}
\usepackage{amsthm}
\usepackage{array} 
\usepackage{booktabs}
\usepackage{mathrsfs}
\usepackage{xcolor}
\usepackage{algorithm}
\usepackage{algorithmicx}
\usepackage{algpseudocode}
\usepackage{enumitem}
\usepackage{longtable}
\usepackage{makecell}

\theoremstyle{thmstyleone}

\theoremstyle{thmstyletwo}

\theoremstyle{thmstylethree}

\title{Motzkin-Straus Optimization on an Entropy-Computing Platform}

\author[1]{\fnm{PoJen} \sur{Wang}}\email{pwang@quantumcomputinginc.com}
\author[1]{\fnm{Sutapa} \sur{Samanta}}
\author[1,2]{\fnm{Yuntai} \sur{Song}}
\author[1]{\fnm{Mohammad-Ali} \sur{Miri}}

\affil*[1]{\orgname{Quantum Computing Inc (QCi)}, \orgaddress{\street{5 Marine View Plaza}, \city{Hoboken}, \postcode{07030}, \state{NJ}, \country{USA}}}

\affil*[2]{\orgname{Department of Physics, University of Illinois Urbana-Champaign}, \orgaddress{\street{1110 West Green Street}, \city{Urbana}, \postcode{61801}, \state{IL}, \country{USA}}}

\def\orcidlogo{}
\begin{document}
\abstract{We introduce a framework for combinatorial optimization using sum-constrained continuous quadratic programs solvable by QCi's Dirac-3S photonic entropy computer. This is enabled by the Motzkin-Straus theorem which provides a powerful bridge between discrete clique problems and optimization over the probability simplex. 
We demonstrate this framework's versatility by solving constraint satisfaction problems, providing extensive benchmarks on the DIMACS suite. The Dirac-3S platform matches or outright leads two independently implemented classical baselines on more than four-fifths of the benchmark instances, reaching the best known solution on nearly all structured graph families, even outperforming both classical solvers on several of the largest instances tested. On the other hand, well-tuned classical continuous optimizers retain an edge only on the hardest planted-clique instances. This work establishes a viable pathway for solving combinatorial optimization problems using natively analog unconventional computing platforms, while positioning entropy computing as a competitive approach for navigating non-convex landscapes and providing rigorous baselines for an emerging computational paradigm.}


\maketitle

\section{Introduction}
\label{sec:introduction}

Combinatorial optimization problems in the NP-hard class pose a formidable barrier to progress in fields ranging from logistics and bioinformatics to artificial intelligence. The Maximum Independent Set (MIS) problem, a canonical example, underscores this challenge, with applications central to resource allocation, scheduling, and network analysis~\cite{garey1979computers}. As the scale of these problems grows, the limitations of traditional computing architectures become increasingly apparent. Exact methods can become computationally prohibitive on large or difficult instances, while heuristics, though practical, surrender the guarantee of optimality.

The transformation of the maximum clique problem into a continuous optimization framework via the Motzkin-Straus theorem \cite{motzkin1965maxima} has opened a rich and challenging field of study (see \cite{marino2024review_maxclique} for a recent review spanning classical, neural-network, and quantum approaches). The historical development of classical solvers for this formulation is a testament to the persistent difficulty of its non-convex optimization landscape. The original 1965 formulation, while theoretically elegant, presents a landscape replete with spurious local optima~\cite{doi:10.1287/moor.22.3.754,beretta2023kkt}, which serve as traps for deterministic, gradient-based algorithms and lead to poor convergence. The computational difficulty is further underscored by recent results showing that even convex standard quadratic programs on the simplex become NP-hard under sparsity constraints~\cite{bomze2025sparse_stqp}. In response, subsequent research introduced sophisticated countermeasures, each with its own inherent trade-offs. Regularization techniques \cite{bomze1999maximum, bomze1997evolution}, for instance, were developed to reshape the objective function and eliminate these non-clique solutions, but this success came at the cost of introducing problem-dependent parameters that require careful and often computationally expensive tuning.

Alternative approaches sought to bypass this issue through complete problem reformulation. The symmetric rank-one nonnegative matrix approximation \cite{belachew2015solving}, for example, establishes a direct one-to-one correspondence between its global optima and the maximum clique, but in doing so, it elevates the problem's complexity to a quartic objective that demands more specialized solvers. Even established, general-purpose algorithms for simplex-constrained optimization face significant hurdles \cite{bomze1999maximum}. The widely used projected gradient descent method, for instance, must be augmented with computationally intensive multi-restart strategies to effectively explore the solution space and escape the numerous local maxima. More recently, parallelized differentiable quadratic formulations have been proposed that incorporate clique-informed penalty terms and momentum-based gradient descent to improve convergence and exploration on the MIS problem, though these methods still rely on running many parallel initializations to compensate for the non-convex landscape~\cite{alkhouri2024dataless}. Generative approaches based on discrete diffusion models have also been applied to MIS and maximum clique, sampling from the Boltzmann distribution of the problem Hamiltonian without requiring explicit likelihood computation~\cite{sanokowski2024diffuco}. Ultimately, the classical treatment of the Motzkin-Straus problem is characterized by a series of sophisticated yet compromised strategies, where overcoming the fundamental challenge of a complex landscape invariably introduces new costs in parameter sensitivity, algorithmic complexity, or computational overhead.

Recent advances in quantum computing have opened promising, albeit challenging, new avenues for tackling these problems~\cite{lucas2014ising,farhi2014qaoa,ebadi2022mis}. However, recent benchmarks on Rydberg-atom hardware suggest that demonstrating a computational advantage for MIS may require scaling to thousands of atoms, as classical methods remain highly competitive on current problem sizes~\cite{cazals2025hard_mis}. On superconducting hardware, quantum-enhanced greedy algorithms that use shallow QAOA circuits to guide classical heuristics have demonstrated scalable MIS solving on graphs larger than the physical qubit count~\cite{wybo2026quantum_greedy}. Trapped-ion platforms have also been explored for MIS through digital-analog counterdiabatic protocols that reduce circuit depth~\cite{kumar2025trapped_ion_mis}. Despite these advances across multiple platforms, a distinct approach has emerged in the form of entropy computing, as realized in QCI's Dirac-3S system~\cite{nguyen2024entropy}. This paradigm marks a departure from mainstream quantum computing's reliance on fragile, isolated quantum systems. Instead, it operates by conditioning a quantum reservoir to stabilize ground states through controlled dissipation and measurement-feedback loops~\cite{verstraete2009dissipation,kraus2008quantum_markov,wiseman2010quantum}. Entropy computing paradigm harnesses quantum effects, such as fluctuations and shot noise, not as sources of error to be suppressed, but as computational resources to escape local minima. This offers a unique and potentially powerful approach for navigating the complex, non-convex landscapes characteristic of NP-hard problems, which is distinct from the algorithmic heuristics like multi-start framework \cite{hungerford2017generalregularizedcontinuousformulation}.

\begin{figure}[htbp]
\centering
\includegraphics[width=1.0\textwidth]{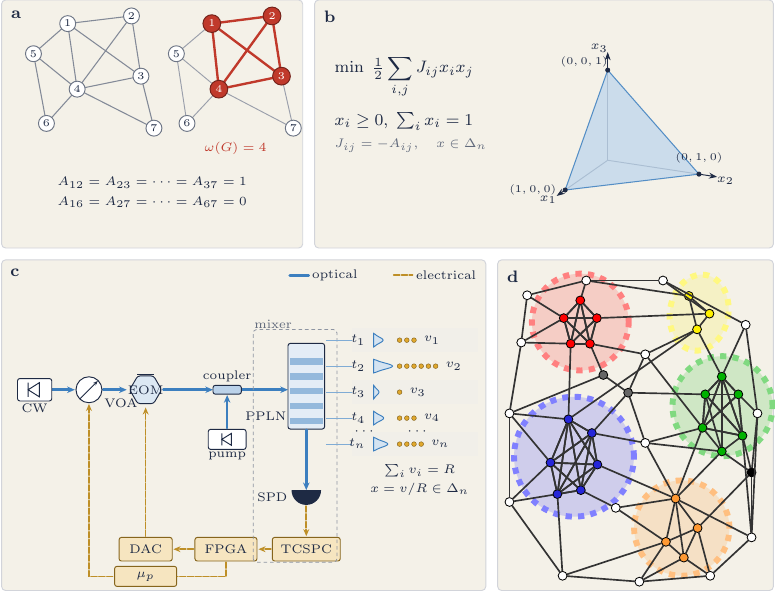}
\caption{\textbf{A schematic illustration of Motzkin-Straus optimization through entropy computing}. (a) The maximum clique problem. (b) The Motzkin-Straus mapping of the maximum clique problem to the corresponding quadratic optimization problem over non-negative variables subject to a sum constraint. (c) The entropy computing architecture utilizing photon number encoding in time bins respecting the non-negativity and sum constraints for faithful Motzkin-Straus optimization. (d) Applications of the maximum clique problem in community detection~\cite{fortunato2010community}.}
\label{fig1}
\end{figure}

This work explores the intersection of classical optimization theory and emerging entropy computing hardware, establishing several significant contributions to the field of quantum-enhanced combinatorial optimization. Our primary contribution lies in the design and implementation of Motzkin-Straus oracles operating on an entropy computer, as illustrated in Fig.~\ref{fig1}. The oracle implementation exploits the natural quadratic structure of the Motzkin-Straus theorem, which provides an exact relationship $\omega(G) = \frac{1}{1-2M^*}$ between the clique number and the optimal value $M^*$ of the continuous optimization problem. By implementing this optimization directly on the entropy computing's Dirac-3S platform, we eliminate the approximation errors and computational overhead associated with discrete-to-continuous problem translations that characterize most quantum approaches to combinatorial optimization. Our experimental methodology provides a rigorous comparative performance analysis through comprehensive benchmarking on the standard DIMACS graph suite~\cite{johnson1996cliques}, employing highly optimized classical implementations including projected gradient descent with Armijo backtracking (Armijo-PGD)~\cite{Armijo1966MinimizationOF,birgin2000spg} and Hexaly~\cite{Hexaly15}, a commercial general-purpose mathematical programming solver, as independent classical baselines for the same unregularized quadratic program. 
%
Among related photonic optimization platforms, time-multiplexed measurement-feedback systems such as coherent Ising machines~\cite{inagaki2016cim,mcmahon2016cim,yamamoto2020cim,yamamura2017cim_quantum_model} share architectural similarities with the Dirac-3S approach. Notably, a recent CIM demonstration solved the MIS problem on dense graphs with up to 40{,}000 nodes, outperforming optimized simulated annealing in time-to-solution for large instances~\cite{takesue2025cim_mis}. The hardware architecture of the Dirac-3S platform and its natural mapping to the Motzkin-Straus formulation are described in detail in Section~\ref{sec:EQC}.
\section{Max-clique Problem}
\subsection{Problem Formulation}
\label{sec:problem_formulation}
The mathematical cornerstone of our work is the Motzkin-Straus theorem~\cite{motzkin1965maxima}, which reveals a profound connection between discrete graph theory and continuous optimization. For a graph $G = (V, E)$ with adjacency matrix $A$, the theorem states:
\begin{equation}
\max_{x \in \Delta_n} \frac{1}{2} x^T A x = \frac{1}{2}\left(1 - \frac{1}{\omega(G)}\right)
\label{eq:motzkin_straus}
\end{equation}
where $\Delta_n = \{x \in \mathbb{R}^n : \sum_{i=1}^n x_i = 1, x_i \geq 0\}$ is the standard probability simplex and $\omega(G)$ is the size of the maximum clique in $G$. For a scaled simplex $\Delta_n^R = \{y \in \mathbb{R}^n : \sum_{i=1}^n y_i = R, y_i \geq 0\}$, Eqn.~\ref{eq:motzkin_straus} is modified to
\begin{equation}
\max_{y \in \Delta_n^R} \frac{1}{2} y^T A y = \frac{R^2}{2}\left(1 - \frac{1}{\omega(G)}\right)
\label{eq:motzkin_straus_scaled}
\end{equation}

This equivalence allows for the creation of oracles that can determine the clique number of a graph, a canonical NP-hard problem, by solving a continuous quadratic program. By leveraging the fundamental relationship $\alpha(G) = \omega(\overline{G})$, where $\alpha(G)$ is the independence number of $G$ and $\overline{G}$ is its complement, we establish a complete algorithmic framework for the Maximum Independent Set problem through optimization on the complement graph.

\subsection{Clique Extraction from Continuous Solutions}
\label{sec:clique_extraction}
Given a graph $G = (V, E)$ with adjacency matrix $A$, we minimize $f= \tfrac12 y^T J y$, with $J=-A$. Let $f^*$ be the minimum, then
\begin{equation}
    f^* = -\frac{R^2}2 \left(1-\frac1{\omega(G)}\right),\qquad 
    \text{equivalently}\qquad
    \omega(G) = \frac1{1+2f^*/R^2}\, .
    \label{eq:compute_omega}
\end{equation}
 Because the formula divides the energy by $R^2$, it is invariant to the choice of $R$.
 
Any feasible solution therefore yields a continuous estimate for $\omega$ which takes non-integer value in general. The continuous estimate of $\omega$ might not be clique at all as there are spurious maximisers supported on non-cliques~\cite{jagota1995feasible}.
We must extract discrete integer value of max-cut from the continuous estimation of $\omega$. We employ the following extraction methods. First, we use the Motzkin-Straus relationship in~\ref{eq:compute_omega} to estimate the clique number directly from the continuous optimum by rounding to closest integer. Then the \emph{support-based} method identifies the support set $S = \{i : x^*_i > \tau\}$ for a threshold $\tau > 0$ and extracts a valid clique from the subgraph induced by $S$. For small supports ($|S| \leq 20$), all subsets are enumerated to find the largest clique. For larger supports, greedy peeling which iteratively removes the vertex with the most non-adjacencies, and greedy building which iteratively adds the heaviest vertex adjacent to all selected vertices are applied, and the best result is retained. All candidate cliques are extended to maximal cliques before reporting. The full algorithmic details are provided in the supplementary information.

\section{Solvers}
\subsection{Dirac-3S}
\label{sec:EQC}
The Dirac-3 platform~\cite{nguyen2024entropy} implements the entropy computing paradigm through a hybrid quantum-classical measurement-feedback system that encodes optimization variables in time-multiplexed photon qudits. Dirac-3S is the second generation Dirac-3 solver with higher variable count. A schematic of Dirac-3S hardware is illustrated in Fig.~\ref{fig1}(c).
The system minimizes a polynomial cost function $E$ over non-negative variables $v_i$~\cite{nguyen2024entropy}:
\begin{multline}
    E = \sum_{i} C_i v_i +
        \sum_{i,j} J_{ij} v_i v_j +
        \sum_{i,j,k} T_{ijk} v_i v_j v_k +\\
        \sum_{i,j,k,l} Q_{ijkl} v_i v_j v_k v_l +
        \sum_{i,j,k,l,m} P_{ijklm} v_i v_j v_k v_l v_m,
\label{eq:dirac_cost}
\end{multline}
where $C_i$, $J_{ij}$, $T_{ijk}$, $Q_{ijkl}$, and $P_{ijklm}$ are fully programmable weight tensors symmetric under permutation of indices, supporting up to fifth-order interactions.
Two constraints arise naturally from the photonic hardware:
(i) \emph{non-negativity}: $v_i \geq 0$ for all $i$, since photon counts are intrinsically non-negative; and
(ii) \emph{sum constraint}: $\sum_i v_i = R$, where $R$ is set by the total photon flux.

The Motzkin-Straus quadratic program maps to the Dirac-3S hardware with exceptional directness.
The maximum clique formulation requires maximizing the quadratic objective $\tfrac{1}{2} x^T A x$ subject to the simplex constraint $x \in \Delta_n$, i.e., $\sum_i x_i = 1$ and $x_i \geq 0$.
Each of these three requirements is natively satisfied by the photonic architecture without any encoding overhead:
\begin{enumerate}[leftmargin=*]
    \item The quadratic objective $x^T A x$ maps directly to the two-body interaction term $\sum_{i,j} J_{ij} v_i v_j$ in Eq.~\eqref{eq:dirac_cost} by setting $J_{ij} = -A_{ij}$ and all higher-order tensors to zero. Here $A_{ij}$ is the adjacency matrix 
    \item The simplex constraint $\sum_i x_i = 1$ is provided by the hardware-enforced photon conservation $\sum_i v_i = R$ with $R = 1$, arising from the fixed total photon flux in the feedback loop.
    \item The non-negativity constraint $x_i \geq 0$ is intrinsic to photon counting. As a result no slack variables or penalty terms are required.
\end{enumerate}
This stands in contrast to approaches based on the Ising Hamiltonian, where encoding the Motzkin-Straus problem would require penalty terms to enforce the simplex constraint, quadratization of any higher-order terms, and minor embedding to handle dense connectivity.
The entropy computing paradigm thus provides a direct physical realization of the mathematical structure of the maximum clique problem, without any translation layer between the optimization formulation and the hardware.

Quantum fluctuations serve as a built-in mechanism for escaping local optima in the non-convex Motzkin-Straus landscape.
The photon counting process follows Poisson statistics, producing a quantum fluctuation coefficient of $1/\sqrt{N}$, where $N$ is the number of photons accumulated per time bin.
Lower mean photon number $\mu$ increases the relative magnitude of shot noise, thereby enhancing stochastic exploration of the solution space.
Empirical evidence from the original Dirac-3 experiments demonstrates that operating in the deep single-photon regime, where the probability of multi-photon events is below $1\%$, yields significantly improved solution quality on non-convex problems~\cite{nguyen2024entropy}.
This noise-driven exploration is functionally analogous to temperature in simulated annealing, but arises directly from quantum measurement physics rather than from an artificial noise schedule.
A trade-off exists: lower photon numbers improve exploration but increase the measurement time required for photon accumulation, creating a tunable balance between solution quality and computational speed.

In its current implementation, the Dirac-3S platform supports up to 9980 independent variables under summation constraints, with configurable sampling parameters \texttt{num\_samples} $\in [1,100]$ and \texttt{relaxation\_schedule} $\in \{1,2,3,4\}$ that control the rate of fluctuation annealing across feedback iterations.
The experiments reported in this work use graphs with up to approximately 4000 vertices, which fall within the comfortable operating range of the device.

\subsection{Classical Solvers}
We selected two classical optimization approaches that represent the current performance frontier while offering distinct computational characteristics: Projected Gradient Descent with Armijo backtracking~\cite{Armijo1966MinimizationOF,birgin2000spg}, and Hexaly~\cite{Hexaly15}. These methods were selected as they represent state-of-the-art techniques for the direct optimization of the Motzkin-Straus quadratic program~\cite{gibbons1996continuous,doi:10.1287/moor.22.3.754}. Other classical approaches to maximum clique and related graph problems include local-search and tabu-based heuristics~\cite{grosso2008heuristics,WU2015693}, parallel clique algorithms~\cite{a6040618}, differentiable quadratic formulations~\cite{alkhouri2024dataless}, and unsupervised graph neural network methods~\cite{karalias2021erdos_neural,acikalin2025x2gnn}. These approaches employ different formulations or search strategies and are not evaluated here. Our objective is to compare Dirac-3S, Armijo-PGD, and Hexaly on the same unregularized Motzkin--Straus quadratic program. The comparison therefore assesses the tested solver configurations on a common formulation, rather than establishing their competitiveness against the strongest specialized maximum-clique algorithms.

\subsubsection{Armijo-PGD}
\label{sec:Armijo}
The Projected Gradient Descent method with Armijo line search provides our primary classical baseline, combining theoretical rigor with robust practical performance across diverse graph structures~\cite{belachew2015solving,birgin2000spg}. The method addresses the non-convex Motzkin-Straus optimization landscape through adaptive step-size selection and strategic multi-restart initialization.

At iterate $x_k \in \Delta_n$ with gradient $g_k = \nabla f(x_k) = Ax_k$ for the maximization problem, the algorithm forms a trial point through gradient step followed by simplex projection:
\begin{equation}
\widetilde{x}(\alpha) = \Pi_{\Delta_n}(x_k + \alpha g_k)
\end{equation}
where $\Pi_{\Delta_n}$ denotes the Euclidean projection onto the probability simplex, computed efficiently in $O(n \log n)$ time using the sorting-based algorithm of Duchi et al.~\cite{duchi2008efficient}.

The Armijo condition provides adaptive step-size selection by accepting step size $\alpha$ when the sufficient decrease inequality~\cite{Armijo1966MinimizationOF,birgin2000spg}
\begin{equation}
f(\widetilde{x}(\alpha)) \geq f(x_k) + c_1 \alpha \nabla f(x_k)^T (\widetilde{x}(\alpha) - x_k)
\end{equation}
holds for fixed parameter $c_1 \in (0,1)$. Starting from initial step size $\alpha_{\text{init}}$, the backtracking procedure reduces the step by factor $\beta \in (0,1)$ until the condition is satisfied. This adaptive mechanism guarantees monotone increase while automatically adjusting to the local geometry of the optimization landscape.

The multi-restart framework employs Dirichlet initialization with concentration parameter $\alpha = 0.5$ to sample diverse starting points across the probability simplex. This initialization strategy addresses the sensitivity to starting points that characterizes non-convex optimization by ensuring comprehensive exploration of promising regions. Our implementation executes multiple independent optimization trajectories and selects the solution achieving the highest objective value.

Our implementation employs 100 independent restarts with maximum 20,000 iterations per restart. Implementation leverages JAX~\cite{jax2018github} for Just-In-Time compilation and automatic differentiation, enabling efficient vectorized execution of multiple restarts. The configuration employs parameters $c_1 = 10^{-4}$, $\beta = 0.5$, and $\alpha_{\text{init}} = 1.0$, chosen based on empirical validation across DIMACS benchmark instances to optimize the trade-off between convergence speed and solution quality.

\subsubsection{Hexaly}
\label{sec:hexaly}
Hexaly is a commercial general-purpose mathematical programming solver built on a portfolio of local-search and direct-search heuristics with exact components. We use Hexaly 15.0~\cite{Hexaly15} to solve the problem instances. We declare $n$ continuous decision variables $y\ge0$, one linear equality constraint $\sum_i y_i = R$, and minimise the objective. The quadratic form is supplied as a black-box external function together with its analytic gradient:
\begin{equation}
  f_{\text{ext}}(y) = y^T J y, \qquad
  \nabla f_{\text{ext}}(y) = 2Jy .
\end{equation}
Supplying the gradient rather than letting Hexaly finite-difference helps it scale well at large number of variables. 

Hexaly solver has a user defined \texttt{time\_limit} parameter. For non-convex problems, Hexaly tends to keep running until the limit expires, irrespective of whether objective function is saturated to a value or not. We set the time limit to be 300 seconds for our experiments.


\section{Results}
\subsection{Benchmark On DIMACS}
We evaluate our approach on standard DIMACS benchmark graphs~\cite{johnson1996cliques} that provide comprehensive coverage of the diverse structural characteristics encountered in real-world maximum clique problems. The Brock series consists of structured graphs with known optimal solutions spanning from brock200-1 through brock800-4, providing systematic scalability analysis across increasing problem sizes while maintaining known ground truth values for precise performance assessment.

The C-fat series encompasses dense clique-rich graphs including c-fat200-1 and c-fat500-10, designed specifically to challenge maximum clique algorithms through high edge density and multiple large cliques that create complex optimization landscapes with numerous competitive solutions. These instances test algorithmic ability to distinguish between near-optimal and optimal solutions in highly connected graphs.

Johnson graphs represent combinatorial designs with well-understood mathematical properties that enable theoretical analysis of algorithmic performance, providing instances where structural characteristics can be leveraged to validate theoretical predictions about optimization behavior. Hamming graphs encode error-correcting code structures that arise naturally in coding theory applications, offering practical relevance beyond academic benchmarking.

Keller graphs provide geometric constructions with high chromatic numbers that challenge algorithms through complex constraint interactions arising from geometric embedding properties.

We evaluated the accuracy of our oracles on a broad selection of 75 DIMACS benchmark graphs, with sizes ranging from 28 to 4000 vertices. The primary goal of this benchmark is to understand the solution quality of the emergent entropy computing approach compared to highly refined classical solvers for the same continuous optimization problem. For this purpose, we compare the Dirac-3S~\ref{sec:EQC} solver against Armijo-PGD~\ref{sec:Armijo} and Hexaly~\ref{sec:hexaly}.

Table~\ref{tab:benchmark_results} reports the results for all 75 graphs. The `Best Known' column lists the accepted optimal clique size from the literature or the best known lower bound, marked $\geq$, where optimality has not been proven. We report the integer clique size extracted from the solutions provided by the Dirac-3S, the classical Armijo-PGD, and the Hexaly oracles.

\scriptsize
\begin{longtable}{lrrrccc}
\caption{Benchmark Results on Selected DIMACS Graphs. Bold indicates the solver that beat others. Star indicates when Dirac-3S solution is the best.}\label{tab:benchmark_results}\\
\toprule
Graph & Vertices & Edges &\thead{Best\\ Known} & Dirac-3S & \thead{Armijo-\\PGD} & Hexaly\\ 
\midrule
C125.9           & 125 & 6,963 &  34 & 34 & 34 & 34\\
C250.9           & 250 & 27,984 &  44 & \bf{44}* & 42 & 42\\
C500.9           & 500 & 112,332 & $\ge$ 57 & \bf{53*} & 52 & 52\\
C1000.9          & 1000 & 450,079 & $\geq 68$  & \bf{63*} & 61 & 62\\
C2000.5          & 2000 & 999,836 & 16  & \bf{15*}  & 14 & 14\\
C2000.9          & 2000 & 1,799,532  & $\geq 80$  & \bf{70*} & 68 & 67\\
C4000.5          & 4000 & 4,000,268  & 18 & \bf{15} & \bf{15} & 12\\
DSJC500.5        & 500 & 62,624 & 13 & 12 & 12 & 12\\
DSJC1000.5       & 1000 & 249,826  & 15 & \bf{14} & 13 & \bf{14}\\
brock200\_1      & 200 & 14,834  & 21 & 20 & 20 & 20\\
brock200\_2      & 200 & 9,876  & 12 & 10 & 10 & 10 \\
brock200\_3      & 200 & 12,048 & 15 & 13 & 13 & 13\\
brock200\_4      & 200 & 13,089  & 17 & \bf{16} & 15 &  \bf{16}\\
brock400\_1      & 400 & 59,723  & 27 & \bf{24} & 23 & \bf{24}\\
brock400\_2      & 400 & 59,786  & 29 & \bf{24} & \bf{24} & 23\\
brock400\_3      & 400 & 59,681 & 31 & \bf{25*} & 23 & 24\\
brock400\_4      & 400 & 59,765 & 33 & \bf{24} & 23 & \bf{24}\\
brock800\_1      & 800 & 207,505 & 23 & 19 & 19 & 19 \\
brock800\_2      & 800 & 208,166  & 24 & \bf{20} & 19 & \bf{20}\\
brock800\_3      & 800 & 207,333 & 25 & \bf{21*} & 19 & 19\\
brock800\_4      & 800 & 207,643 & 26 & \bf{20*} & 19 & 19\\
c-fat200-1       & 200 & 1,534  & 12 & 12 & 12 & 12\\
c-fat200-2       & 200 & 3,235  & 24 & 24 & 24 & 24\\
c-fat200-5       & 200 & 8,473  & 58 & 58 & 58 & 58\\
c-fat500-1       & 500 & 4,459  & 14 & 14 & 14 & 14\\
c-fat500-2       & 500 & 9,139  & 26 & 26 & 26 & 26\\
c-fat500-5       & 500 & 23,191  & 64 & 64 & 64 & 64\\
c-fat500-10      & 500 & 46,627  & 126 & 126 & 126 & 126\\
gen200-p0.9-44   & 200 & 17,910  & 44 & 38 & \bf{39} & 38\\
gen200-p0.9-55   & 200 & 17,910 & 55 & 42 & \bf{47} & 40\\
gen400-p0.9-55   & 400 & 71,820 & 55 & \bf{51*} & 50 & 50\\
gen400-p0.9-65   & 400 & 71,820 & 65 & \bf{54*} & 49 & 51\\
gen400-p0.9-75   & 400 & 71,820  & 75 & 52 & \bf{53} & 44 \\
hamming6-2       & 64 & 1,824  & 32 & 32 & 32 & 32\\
hamming6-4       & 64 & 704  & 4 & 4 & 4 & 4\\
hamming8-2       & 256 & 31,616 & 128 & 128 & 128 & 128\\
hamming8-4       & 256 & 20,864 &  16 & 16 & 16 & 16\\
hamming10-2      & 1024& 518,656  & 512 &  512 & 512 & 512\\
hamming10-4      & 1024 & 434,176 & $\geq 40$ & \bf{36*} & 35 & 33\\
johnson16-2-4    & 120 & 5,460 & 8 & 8 & 8 & 8\\
johnson32-2-4    & 496 & 107,880  & $\geq$ 16 & 16 & 16 & 16\\
johnson8-2-4     & 28 & 210  & 4 & 4 & 4 & 4\\
johnson8-4-4     & 70 & 1,855  & 14 & 14 & 14 & 14\\
keller4          & 171 & 9,435  & 11 & 9 & 9 & 9\\
keller5          & 776 & 225,990 & 27 & \bf{19*} & 18 & 16\\
keller6          & 3361 & 4,619,898 & 59 & \bf{35*} & 33 & 31\\
p\_hat300-1    & 300 & 10,933  & 8 & 8 & 8 & 8\\
p\_hat300-2    & 300 & 21,928  & 25 & 25 & 25 & 25\\
p\_hat300-3    & 300 & 33,390  & 36 & \bf{36}* & 34 & 35\\
p\_hat500-1    & 500 & 31,569 & 9 & 9 & 9 & 9\\
p\_hat500-2    & 500 & 62,946  & 36 & 36 & 36 & 36\\
p\_hat500-3    & 500 & 93,800  & 50 & 49 & \bf{50} & 49\\
p\_hat700-1    & 700 & 60,999  & 11 & \bf{11} & \bf{11} & 9\\
p\_hat700-2    & 700 & 121,728  & 44 & 44 & 44 & 44\\
p\_hat700-3    & 700 & 183,010  & 62 & 62 & 62 & 62\\
p\_hat1000-1   & 1000 & 122,253 & 10 & \bf{10} & \bf{10} & 9\\
p\_hat1000-2   & 1000 & 244,799 & 46 & \bf{46*} & 45 & 45\\
p\_hat1000-3   & 1000 & 371,746 & 68 & 64 & \bf{66} & 64\\
p\_hat1500-1   & 1500 & 284,923  & 12 & \bf{11} & \bf{11} & 10\\
p\_hat1500-2   & 1500 & 568,960 & 65 & \bf{65} & \bf{65} & 63\\
p\_hat1500-3   & 1500 & 847,244 & 94 & 92 & 92 & 92\\
sanr200\_0.7   & 200 & 13,868 & 18 & 18 & 18 & 18\\
sanr200\_0.9   & 200 & 17,863 &  42 & 41 & 41 & 41\\
sanr400\_0.5   & 400 & 39,984 & 13 & \bf{13*} & 12 & 12\\
sanr400\_0.7   & 400 & 55,869 & 21 & \bf{21*} & 20 & 20\\
san200\_0.7\_1 & 200 & 13,930 & 30 & 16 & \bf{23} & 16\\
san200\_0.7\_2 & 200 & 13,930 & 18 & 12 & \bf{15} & 12\\
san200\_0.9\_1 & 200 & 17,910 & 70 &  47 & \bf{50} & 47\\
san200\_0.9\_2 & 200 & 17,910 & 60 & 42 & \bf{60} & \bf{60}\\
san200\_0.9\_3 & 200 & 17,910 &  44 & \bf{36} & \bf{36} & 35\\
san400\_0.5\_1 & 400 & 39,900 &  13 & 7 & \bf{13} & 7\\
san400\_0.7\_1 & 400 & 55,860 &  40 & 20 & \bf{40} & 20\\
san400\_0.7\_2 & 400 & 55,860 &  30 & 15 & \bf{30} & 15\\
san400\_0.7\_3 & 400 & 55,860 &  22 & 14 & \bf{16} & 12\\
san400\_0.9\_1 & 400 & 71,820 &  100 & 100 & 100 & 100\\
\bottomrule
\end{longtable}
\normalsize
Across the 75 graphs, all three solvers return the same clique size on 33 instances, and on 25 of those the common value is the best known one. On a further 13 instances two of the three solvers tie for the highest value. Dirac-3S and Armijo-PGD tie ahead of Hexaly on 7, Dirac-3S and Hexaly tie ahead of Armijo-PGD on 5, and Armijo-PGD and Hexaly tie ahead of Dirac-3S on 1. So on 46 of the 75 graphs no single solver stands alone at the top. On the remaining 29 graphs Dirac-3S achieves best on 17, Armijo-PGD on 12, and Hexaly on none. Altogether, Dirac-3S matches or exceeds both classical solvers on 62 of the 75 graphs, i.e., on more than four-fifths of the benchmark. Where Dirac-3S is the sole leader, its lead is only 1-3 vertices. Armijo-PGD's leads are likewise 1-2 vertices outside the planted-clique families, apart from gen200-p0.9-55 (5 vertices), but on several \texttt{san} instances it recovers the planted clique outright (san400\_0.5\_1, san400\_0.7\_1, san400\_0.7\_2) with margins of up to 20 vertices. Dirac-3S reaches the best known value on 33 of the 75 graphs, Armijo-PGD on 33, and Hexaly on 26.

Among the 14 graphs with at least 1000 vertices, ties account for 7 of the 14. All three solvers agree on hamming10-2 and p\_hat1500-3, Dirac-3S and Armijo-PGD tie ahead of Hexaly on C4000.5, p\_hat1000-1, p\_hat1500-1 and p\_hat1500-2, and Dirac-3S and Hexaly tie ahead of Armijo-PGD on DSJC1000.5. Dirac-3S is the sole leader on the remaining large graphs C1000.9, C2000.5, C2000.9, hamming10-4, keller6 and p\_hat1000-2, and Armijo-PGD is the sole leader only on p\_hat1000-3. Dirac-3S is therefore at or above both classical solvers on 13 of the 14 large graphs. Taken together, an emerging hardware architecture is competitive with two independently implemented classical solvers on a substantial part of the benchmark.
On the Brock series all solvers struggle, which is consistent with this difficulty being a property of the Motzkin-Straus optimization landscape on these graphs rather than an artifact specific to the entropy computing hardware or to the classical algorithm implementation. The planted-clique \texttt{gen} and \texttt{san} instances are also hard for all three solvers, but here Armijo-PGD's many independent restarts give it a clear advantage over Dirac-3S and Hexaly.

\begin{figure}[ht]
    \centering
    \includegraphics[width=0.8\linewidth]{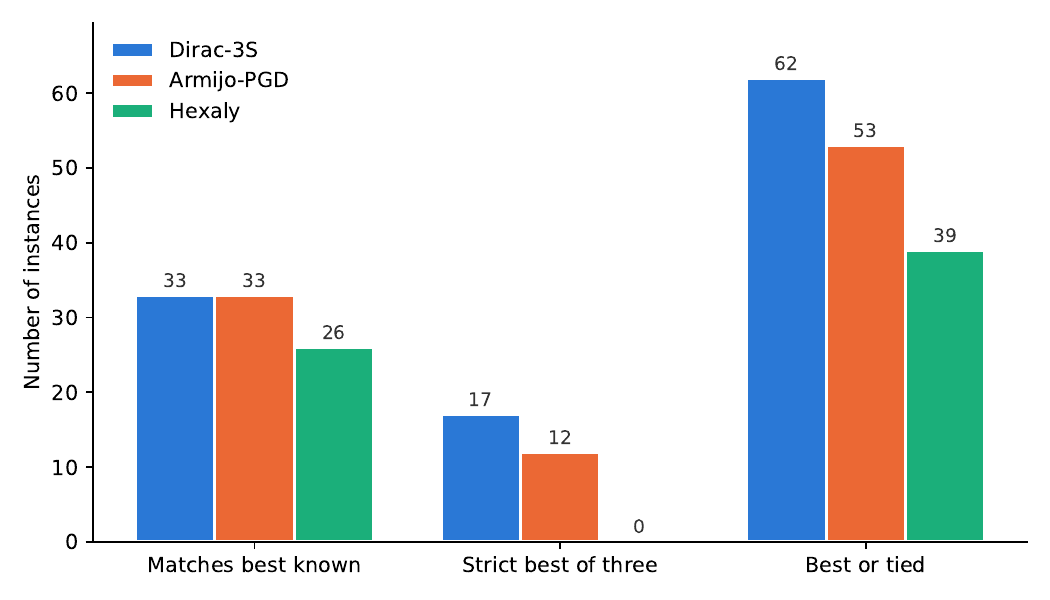}
    \caption{\textbf{Summary of solver performance on the 75 DIMACS benchmark graphs.} For each solver, the bars show the number of instances on which it (i) \emph{matches best known}, i.e.\ returns a clique size equal to the best known value (or best known lower bound) in Table~\ref{tab:benchmark_results}; (ii) is the \emph{strict best of three}, i.e.\ returns a clique strictly larger than those of both other solvers; and (iii) is \emph{best or tied}, i.e.\ returns a clique at least as large as those of both other solvers, including ties. The counts in each category are taken directly from Table~\ref{tab:benchmark_results}.}
    \label{fig:solver_comparison}
\end{figure}

Figure~\ref{fig:solver_comparison} summarizes these head-to-head counts over the full benchmark. The three groups of bars measure progressively relative notions of success. The first counts agreement with the literature, the second counts outright wins, and the third counts instances on which a solver is not beaten by either competitor. Dirac-3S and Armijo-PGD reach the best known value equally often, so the absolute measure alone does not separate them. The difference appears in the relative measures. Dirac-3S is the sole leader more often than Armijo-PGD (17 versus 12), and it is at or above both competitors on 62 instances, compared with 53 for Armijo-PGD and 39 for Hexaly. 

\subsection{Energy Histogram of Specific Examples}
\begin{figure}[htbp]
\centering
\includegraphics[width=1.0\textwidth]{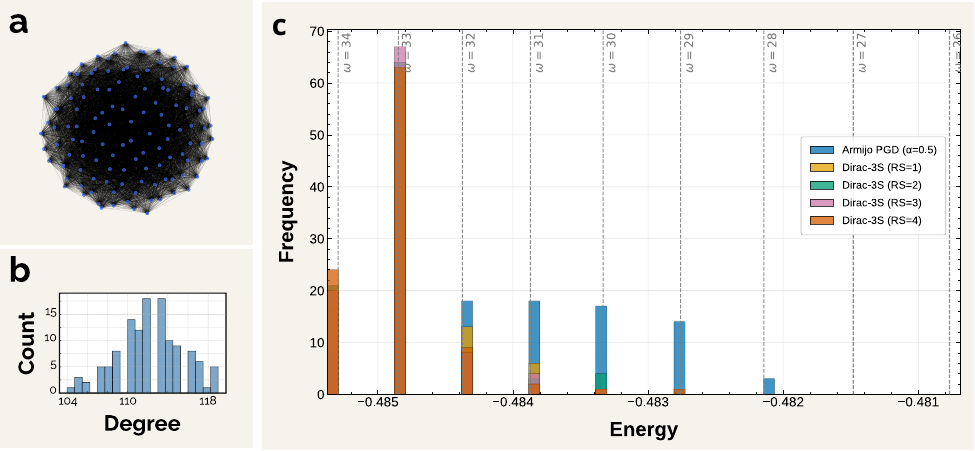}
\caption{\textbf{Energy distribution of Dirac-3S and Armijo-PGD solutions for the DIMACS graph C125.9}. (a) Visualization of the C125.9 graph. (b) Degree distribution of the C125.9 graph. (c) Histogram of the Motzkin-Straus energy values returned across all runs, colored by solver: Armijo-PGD with initial points sampled from a Dirichlet distribution with $\alpha=0.5$, and Dirac-3S at each of its four \texttt{relaxation\_schedule} (RS) settings. Vertical dashed lines mark the theoretical energy $-\frac{1}{2}\left(1-\frac{1}{l}\right)$ predicted by the Motzkin-Straus formula for a clique of size $l$, with the corresponding $l$ labeled above each line; the best known clique number for C125.9 is 34.}
\label{fig:c125_9_dist}
\end{figure}

\begin{figure}[htbp]
\centering
\includegraphics[width=1.0\textwidth]{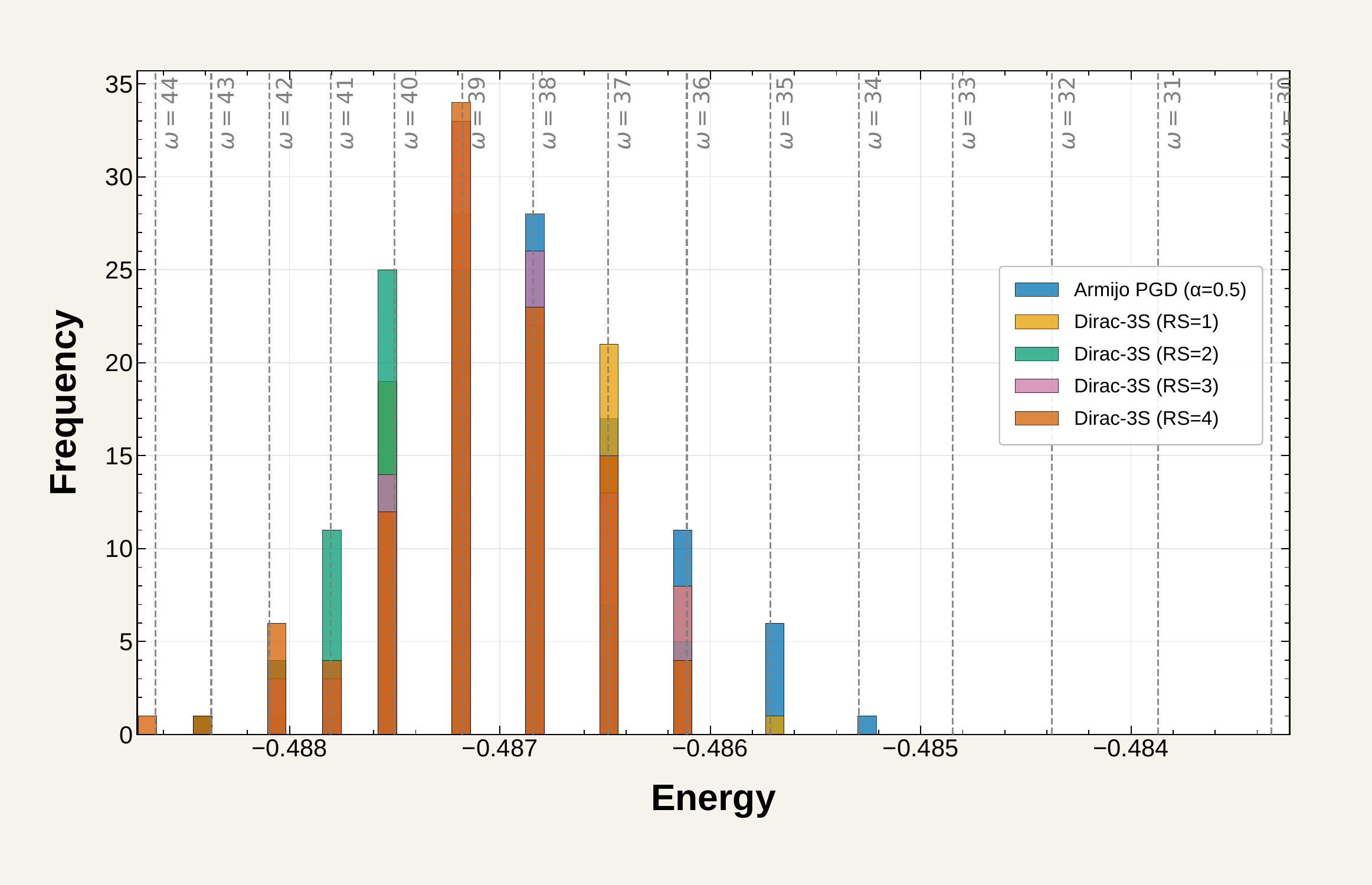}
\caption{\textbf{Energy distribution of Dirac-3S and Armijo-PGD solutions for the DIMACS graph C250.9}. Histogram of the Motzkin-Straus energy values returned across all runs, colored by solver: Armijo-PGD with initial points sampled from a Dirichlet distribution with $\alpha=0.5$, and Dirac-3S at each of its four \texttt{relaxation\_schedule} (RS) settings. Vertical dashed lines mark the theoretical energy $-\frac{1}{2}\left(1-\frac{1}{l}\right)$ predicted by the Motzkin-Straus formula for a clique of size $l$, with the corresponding $l$ labeled above each line; the best known clique number for C250.9 is 44.}
\label{fig:c250_9_dist}
\end{figure}

Figures~\ref{fig:c125_9_dist} and~\ref{fig:c250_9_dist} reveal a qualitative difference in how the two solvers' solutions are distributed relative to the optimum. On the easy instance C125.9 (Fig.~\ref{fig:c125_9_dist}), Dirac-3S's energy distribution is sharply concentrated at the theoretical energies corresponding to $l=33$ and the best known clique size $l=34$, with the large majority of runs across all relaxation-schedule settings landing on one of these two values. Armijo-PGD's distribution is comparatively spread out, with a non-trivial fraction of restarts terminating at suboptimal energies corresponding to $l=28$ through $34$, so that a larger number of independent restarts is needed before the optimum is reliably recovered. The same pattern persists, and is more pronounced, on the harder C250.9 instance (Fig.~\ref{fig:c250_9_dist}): Dirac-3S's mass is shifted toward the higher clique sizes near the optimum, with samples reaching all the way to the best known value $l=44$, whereas Armijo-PGD's distribution remains concentrated at the comparatively lower $l=34$ -- $42$ range and does not reach $l=44$ in the runs shown. This indicates that the entropy computing hardware's measurement-feedback dynamics explore the Motzkin-Straus landscape in a way that concentrates solutions near the optimum, whereas Armijo-PGD's gradient-based restarts are more broadly distributed across suboptimal basins and consequently require a larger sample size to reach the same solution quality.

Table~\ref{tab:solver_comparison} compares three solver paradigms on representative DIMACS instances spanning easy (C125.9), moderate (C250.9), and hard (brock400\_2) difficulty levels. We run Dirac-3S for 100 \texttt{num\_samples}, and report total time to run those 100 samples. Hexaly is run with its fixed 300-second \texttt{time\_limit}, which it consumes in full regardless of instance difficulty. The results reveal distinct performance profiles: all three solvers reach the optimum on the easy instance, C125.9; on the moderate instance, C250.9, only Dirac-3S reaches the optimal 44, while Armijo-PGD and Hexaly fall short; and on the hard instance, brock400\_2, none of the three reaches the optimal 29.

\begin{table}[htbp]
\centering
\caption{Comparative Solver Performance on Representative DIMACS Instances}
\label{tab:solver_comparison}
\begin{tabular}{llccc}
\toprule
\textbf{Solver} & \textbf{Graph} & \textbf{Clique Found / Best} & \textbf{Ratio} & \textbf{Avg.\ Time (s)}  \\
\midrule
Armijo-PGD (GPU) & C125.9 & 34 / 34 & 100\% & $\sim$15  \\
Dirac-3S       & C125.9 & 34 / 34 & 100\% & $\sim$65  \\
Hexaly          & C125.9 & 34 / 34 & 100\% & $\sim$300  \\
\midrule
Armijo-PGD (GPU) & C250.9 & 42 / 44 & 95.5\% & $\sim$10  \\
Dirac-3S       & C250.9 & 44 / 44 & 100\% & $\sim$120  \\
Hexaly          & C250.9 & 42 / 44 & 95.5\% & $\sim$300  \\
\midrule
Armijo-PGD (GPU) & brock400\_2 & 24 / 29 & 82.8\% & $\sim$150  \\
Dirac-3S       & brock400\_2 & 24 / 29 & 82.8\% & $\sim$200  \\
Hexaly          & brock400\_2 & 23 / 29 & 79.3\% & $\sim$300  \\
\bottomrule
\end{tabular}
\end{table}

\section{Discussion}

\subsection{Graph Family Analysis}
Different graph families exhibit varying oracle performance characteristics that reflect the underlying structure of the optimization landscape. Approximation ratios are computed against the best known clique size or best known lower bound where marked $\geq$ in Table~\ref{tab:benchmark_results}, and are summarized by family in Table~\ref{tab:family_summary}. Structured instances (Hamming, Johnson, c-fat) and the \texttt{p\_hat} family are solved essentially exactly by all three solvers, with average ratios of 95 - 99\%. Performance degrades on graphs with less exploitable structure. For example the dense C-series and the \texttt{sanr}/DSJC/Keller group are solved to 85 - 93\% on average. The Brock series, constructed specifically to hide its maximum clique among low-degree vertices, drops further to 81 - 84\%. The planted-clique \texttt{gen} and \texttt{san} families are the hardest for Dirac-3S and Hexaly, averaging only 71\%, whereas Armijo-PGD reaches 86.5\% on this family. Dirac-3S attains the highest average ratio in five of the six families, with Armijo-PGD and Hexaly trailing by small margins. The exception is the planted-clique family, where Armijo-PGD leads by a wide margin; this single family is enough to give Armijo-PGD the highest average ratio over all instances (91.2\% versus 89.2\% for Dirac-3S and 87.4\% for Hexaly). Apart from this, the ranking of family difficulty is broadly consistent across the three mechanistically unrelated solvers, which supports the conclusion that this difficulty gradient largely reflects the structure of the Motzkin-Straus landscape itself rather than a weakness specific to any one solver.
\begin{table}[htbp]
\centering
\caption{Average Approximation Ratio by Graph Family. Ratios are computed against the best known clique size (lower bound where marked $\geq$ in Table~\ref{tab:benchmark_results}).}
\label{tab:family_summary}
\begin{tabular}{lcccc}
\toprule
\textbf{Graph Family} & \textbf{Instances} & \textbf{\thead{Dirac-3S \\ Avg.\ Ratio}} & \textbf{\thead{Armijo-PGD \\ Avg.\ Ratio}} & \textbf{\thead{Hexaly \\ Avg.\ Ratio}} \\
\midrule
Structured (Hamming, Johnson, c-fat) & 17 & 99.4\% & 99.3\% & 99.0\% \\
Dense random (C-series) & 7 & 92.9\% & 90.3\% & 88.0\% \\
Brock (200--800) & 12 & 84.3\% & 81.3\% & 82.7\% \\
Planted clique (gen, san) & 15 & 71.0\% & 86.5\% & 70.8\% \\
p\_hat & 15 & 98.8\% & 98.6\% & 95.8\% \\
Random (sanr, DSJC, keller) & 9 & 88.3\% & 85.4\% & 84.9\% \\
\midrule
All instances & 75 & 89.2\% & 91.2\% & 87.4\% \\
\bottomrule
\end{tabular}
\end{table}

\subsection{Interpreting Benchmark Results Through Platform Characteristics}
\label{sec:quantum_classical_positioning}

The benchmark results can be interpreted through the lens of the solver's property. Classical approaches handle the non-convex Motzkin-Straus landscape through regularization~\cite{bomze1999maximum}, alternative formulations~\cite{belachew2015solving}, population-level dynamics~\cite{pelillo1999replicator,bomze2002annealed}, or general-purpose local-search portfolios such as Hexaly, each introducing complexity through parameter tuning or convergence trade-offs. The Dirac-3S platform instead leverages quantum shot noise as an exploration mechanism, providing a physically distinct alternative to multi-restart strategies. Whether this constitutes a qualitative advantage or is better understood as a hardware-implemented form of stochastic optimization remains an open question~\cite{kadowaki1998quantum_annealing}. Recent work has shown that for certain photonic approaches to graph problems, efficient classical algorithms can replicate quantum sampling distributions~\cite{oh2024gbs_classical}, reinforcing the importance of benchmarking against strong classical baselines -- a role that both Armijo-PGD and Hexaly serve in this study, since they are built on entirely different optimization principles (multi-restart projected gradient ascent versus a black-box heuristic portfolio) yet converge on similar qualitative conclusions about which graph families are hard.

Our results provide some preliminary evidence for this view. Dirac-3S's measurement-feedback dynamics produce diverse solution candidates without explicit multi-start procedures, and the platform is the outright best of the three solvers on a handful of instances that are not confined to small graphs, including six with at least 1000 vertices (Table~\ref{tab:benchmark_results}). On these larger graphs, Dirac-3S is at or above the level of the other two solvers on 13 of the 14 instances, more often than either classical baseline. This asymmetry suggests that the platform's exploration dynamics may complement classical gradient-based methods on some large instances in a way that a third, unrelated classical heuristic does not reproduce. The margins involved are typically only one or two vertices, however, so this evidence is suggestive rather than conclusive, and part of the apparent advantage may simply reflect the generic benefit of consulting a second, independent solver. Characterizing when and why the platform's physical dynamics offer practical benefits over gradient-based methods, particularly near energy barriers, is a key direction for future work.

\subsection{Assessment and Practical Considerations}
Our benchmarks show that the Dirac-3S platform already matches or leads two independently implemented classical solvers -- Armijo-PGD and Hexaly -- on more than four-fifths of the DIMACS instances and attains the highest average approximation ratio in five of six graph families. The remaining gap in raw solution quality is confined mainly to the planted-clique instances, which places Dirac-3S's overall average approximation ratio between Armijo-PGD's and Hexaly's (Table~\ref{tab:family_summary}).

Compared to quantum annealing approaches, the entropy computing architecture offers a significant practical advantage for dense graph problems: the Motzkin-Straus quadratic program is implemented directly on the hardware without embedding overhead. On quantum annealers such as D-Wave, embedding dense graph structures onto sparse hardware connectivity graphs introduces substantial overhead that degrades solution quality~\cite{gherardi2024dwave_maxclique}. The Dirac-3S platform avoids this entirely by satisfying the simplex constraint at the hardware level (see Section~\ref{sec:EQC}).

Furthermore, as a near-term technology deployed on existing hardware rather than future fault-tolerant systems, entropy computing provides a concrete platform for iterative benchmarking and improvement. This positions it as complementary to gate-based quantum algorithms, which require extensive error correction infrastructure that may take years to mature.

\subsection{Current Limitations and a Roadmap for the Future}
The main limitations of this study lie in the method and the benchmark rather than in the platform. Solution quality is weakest on the planted-clique \texttt{gen} and \texttt{san} instances and on the Brock graphs, where the Motzkin-Straus landscape hides the maximum clique among many competing local optima. The comparison is also restricted to the unregularized quadratic formulation and to two classical baselines (Armijo-PGD and Hexaly), and does not include specialized clique heuristics. Finally, each run currently yields a single clique extracted from the highest-weight solution. Future work should therefore focus on hybrid quantum-classical algorithms.

The true power of this paradigm may lie in intelligent hybrid algorithms that combine the complementary strengths of quantum and classical approaches. One compelling direction involves using the Dirac-3S entropy computing oracle to quickly identify promising regions of the search space through natural exploration mechanisms, followed by refinement using classical solvers like projected gradient descent that excel at local optimization. A similar hybrid philosophy has been demonstrated on neutral-atom hardware, where a quantum-informed reduction algorithm uses the quantum processor to resolve the hard core of the graph when classical kernelization stalls, achieving exact solutions on instances where standalone quantum approaches fail entirely~\cite{schuetz2025qredumis}. Developing more sophisticated techniques for extracting and interpreting the solution distributions from quantum measurements represents a rich area for future research, potentially enabling access to multiple high-quality solutions from single quantum computations rather than extracting only the highest-weight solution.

\section{Conclusion}
We have presented a quantum-classical hybrid framework that leverages the Motzkin-Straus theorem to map the NP-hard maximum clique problem into a continuous quadratic program natively suited to QCI's Dirac-3S entropy computer. The key insight is that the Motzkin-Straus formulation's three requirements -- a quadratic objective, simplex constraint, and non-negativity -- are all satisfied at the hardware level by the photonic architecture, eliminating the encoding overhead that characterizes other quantum approaches. Through comprehensive benchmarking on 75 DIMACS instances against two independently implemented classical baselines -- Armijo-PGD and the commercial solver Hexaly -- we demonstrated that the Dirac-3S platform is competitive with both classical baselines overall and matches or leads them on the majority of instances, while well-tuned classical solvers retain an advantage on the hardest planted-clique benchmarks (Table~\ref{tab:family_summary}). 

This work should be viewed as a foundational experiment establishing that entropy computing can be meaningfully applied to canonical NP-hard problems. The main limitation which is the reduced accuracy on planted-clique instances, is a matter of algorithm design rather than a fundamental barrier. Hybrid quantum-classical algorithms offer a clear pathway toward closing the gap with classical solvers and ultimately demonstrating practical quantum advantage for combinatorial optimization.

\backmatter

\bmhead{Acknowledgements}

The authors would like to thank Lac Nguyen and other scientists at Quantum Computing Inc for their valuable discussions and engineering support on the Dirac-3S platform.



\bibliography{references}


\begin{thebibliography}{47}
\ifx \bisbn   \undefined \def \bisbn  #1{ISBN #1}\fi
\ifx \binits  \undefined \def \binits#1{#1}\fi
\ifx \bauthor  \undefined \def \bauthor#1{#1}\fi
\ifx \batitle  \undefined \def \batitle#1{#1}\fi
\ifx \bjtitle  \undefined \def \bjtitle#1{#1}\fi
\ifx \bvolume  \undefined \def \bvolume#1{\textbf{#1}}\fi
\ifx \byear  \undefined \def \byear#1{#1}\fi
\ifx \bissue  \undefined \def \bissue#1{#1}\fi
\ifx \bfpage  \undefined \def \bfpage#1{#1}\fi
\ifx \blpage  \undefined \def \blpage #1{#1}\fi
\ifx \burl  \undefined \def \burl#1{\textsf{#1}}\fi
\ifx \doiurl  \undefined \def \doiurl#1{\url{https://doi.org/#1}}\fi
\ifx \betal  \undefined \def \betal{\textit{et al.}}\fi
\ifx \binstitute  \undefined \def \binstitute#1{#1}\fi
\ifx \binstitutionaled  \undefined \def \binstitutionaled#1{#1}\fi
\ifx \bctitle  \undefined \def \bctitle#1{#1}\fi
\ifx \beditor  \undefined \def \beditor#1{#1}\fi
\ifx \bpublisher  \undefined \def \bpublisher#1{#1}\fi
\ifx \bbtitle  \undefined \def \bbtitle#1{#1}\fi
\ifx \bedition  \undefined \def \bedition#1{#1}\fi
\ifx \bseriesno  \undefined \def \bseriesno#1{#1}\fi
\ifx \blocation  \undefined \def \blocation#1{#1}\fi
\ifx \bsertitle  \undefined \def \bsertitle#1{#1}\fi
\ifx \bsnm \undefined \def \bsnm#1{#1}\fi
\ifx \bsuffix \undefined \def \bsuffix#1{#1}\fi
\ifx \bparticle \undefined \def \bparticle#1{#1}\fi
\ifx \barticle \undefined \def \barticle#1{#1}\fi
\bibcommenthead
\ifx \bconfdate \undefined \def \bconfdate #1{#1}\fi
\ifx \botherref \undefined \def \botherref #1{#1}\fi
\ifx \url \undefined \def \url#1{\textsf{#1}}\fi
\ifx \bchapter \undefined \def \bchapter#1{#1}\fi
\ifx \bbook \undefined \def \bbook#1{#1}\fi
\ifx \bcomment \undefined \def \bcomment#1{#1}\fi
\ifx \oauthor \undefined \def \oauthor#1{#1}\fi
\ifx \citeauthoryear \undefined \def \citeauthoryear#1{#1}\fi
\ifx \endbibitem  \undefined \def \endbibitem {}\fi
\ifx \bconflocation  \undefined \def \bconflocation#1{#1}\fi
\ifx \arxivurl  \undefined \def \arxivurl#1{\textsf{#1}}\fi
\csname PreBibitemsHook\endcsname

\bibitem[\protect\citeauthoryear{Garey and Johnson}{1979}]{garey1979computers}
\begin{bbook}
\bauthor{\bsnm{Garey}, \binits{M.R.}},
\bauthor{\bsnm{Johnson}, \binits{D.S.}}:
\bbtitle{Computers and Intractability: A Guide to the Theory of
  NP-Completeness}.
\bpublisher{W.H. Freeman and Company},
\blocation{New York}
(\byear{1979})
\end{bbook}
\endbibitem

\bibitem[\protect\citeauthoryear{Motzkin and Straus}{1965}]{motzkin1965maxima}
\begin{barticle}
\bauthor{\bsnm{Motzkin}, \binits{T.S.}},
\bauthor{\bsnm{Straus}, \binits{E.G.}}:
\batitle{Maxima for graphs and a new proof of a theorem of tur\'an}.
\bjtitle{Canadian Journal of Mathematics}
\bvolume{17},
\bfpage{533}--\blpage{540}
(\byear{1965})
\end{barticle}
\endbibitem

\bibitem[\protect\citeauthoryear{Marino
  et~al.}{2024}]{marino2024review_maxclique}
\begin{botherref}
\oauthor{\bsnm{Marino}, \binits{R.}},
\oauthor{\bsnm{Buffoni}, \binits{L.}},
\oauthor{\bsnm{Zavalnij}, \binits{B.}}:
A Short Review on Novel Approaches for Maximum Clique Problem: from Classical
  algorithms to Graph Neural Networks and Quantum algorithms
(2024)
\end{botherref}
\endbibitem

\bibitem[\protect\citeauthoryear{Gibbons
  et~al.}{1997}]{doi:10.1287/moor.22.3.754}
\begin{barticle}
\bauthor{\bsnm{Gibbons}, \binits{L.E.}},
\bauthor{\bsnm{Hearn}, \binits{D.W.}},
\bauthor{\bsnm{Pardalos}, \binits{P.M.}},
\bauthor{\bsnm{Ramana}, \binits{M.V.}}:
\batitle{Continuous characterizations of the maximum clique problem}.
\bjtitle{Mathematics of Operations Research}
\bvolume{22}(\bissue{3}),
\bfpage{754}--\blpage{768}
(\byear{1997})
\doiurl{10.1287/moor.22.3.754}
\end{barticle}
\endbibitem

\bibitem[\protect\citeauthoryear{Beretta et~al.}{2025}]{beretta2023kkt}
\begin{barticle}
\bauthor{\bsnm{Beretta}, \binits{G.}},
\bauthor{\bsnm{Torcinovich}, \binits{A.}},
\bauthor{\bsnm{Pelillo}, \binits{M.}}:
\batitle{On generalized {KKT} points for the {Motzkin-Straus} program}.
\bjtitle{Journal of Global Optimization}
\bvolume{91}(\bissue{3}),
\bfpage{535}--\blpage{557}
(\byear{2025})
\end{barticle}
\endbibitem

\bibitem[\protect\citeauthoryear{Bomze et~al.}{2025}]{bomze2025sparse_stqp}
\begin{botherref}
\oauthor{\bsnm{Bomze}, \binits{I.M.}},
\oauthor{\bsnm{Peng}, \binits{B.}},
\oauthor{\bsnm{Qiu}, \binits{Y.}},
\oauthor{\bsnm{Y{\i}ld{\i}r{\i}m}, \binits{E.A.}}:
On tractable convex relaxations of standard quadratic optimization problems
  under sparsity constraints.
Journal of Optimization Theory and Applications
\textbf{204}(3)
(2025)
\end{botherref}
\endbibitem

\bibitem[\protect\citeauthoryear{Bomze et~al.}{1999}]{bomze1999maximum}
\begin{barticle}
\bauthor{\bsnm{Bomze}, \binits{I.M.}},
\bauthor{\bsnm{Budinich}, \binits{M.}},
\bauthor{\bsnm{Pardalos}, \binits{P.M.}},
\bauthor{\bsnm{Pelillo}, \binits{M.}}:
\batitle{Maximum weight cliques}.
\bjtitle{Handbook of combinatorial optimization}
\bvolume{4},
\bfpage{1}--\blpage{74}
(\byear{1999})
\end{barticle}
\endbibitem

\bibitem[\protect\citeauthoryear{Bomze}{1997}]{bomze1997evolution}
\begin{barticle}
\bauthor{\bsnm{Bomze}, \binits{I.M.}}:
\batitle{Evolution towards perfection}.
\bjtitle{Games and economic behavior}
\bvolume{20}(\bissue{2}),
\bfpage{227}--\blpage{249}
(\byear{1997})
\end{barticle}
\endbibitem

\bibitem[\protect\citeauthoryear{Belachew and
  Gillis}{2015}]{belachew2015solving}
\begin{botherref}
\oauthor{\bsnm{Belachew}, \binits{M.T.}},
\oauthor{\bsnm{Gillis}, \binits{N.}}:
Solving the Maximum Clique Problem with Symmetric Rank-One Nonnegative Matrix
  Approximation
(2015)
\end{botherref}
\endbibitem

\bibitem[\protect\citeauthoryear{Alkhouri et~al.}{2025}]{alkhouri2024dataless}
\begin{bchapter}
\bauthor{\bsnm{Alkhouri}, \binits{I.}},
\bauthor{\bsnm{Le~Denmat}, \binits{C.}},
\bauthor{\bsnm{Li}, \binits{Y.}},
\bauthor{\bsnm{Yu}, \binits{C.}},
\bauthor{\bsnm{Liu}, \binits{J.}},
\bauthor{\bsnm{Wang}, \binits{R.}},
\bauthor{\bsnm{Velasquez}, \binits{A.}}:
\bctitle{Differentiable quadratic programming layers for maximum independent
  set}.
In: \bbtitle{International Conference on Learning Representations (ICLR)}
(\byear{2025}).
\bcomment{arXiv:2406.19532}
\end{bchapter}
\endbibitem

\bibitem[\protect\citeauthoryear{Sanokowski
  et~al.}{2024}]{sanokowski2024diffuco}
\begin{bchapter}
\bauthor{\bsnm{Sanokowski}, \binits{S.}},
\bauthor{\bsnm{Hochreiter}, \binits{S.}},
\bauthor{\bsnm{Lehner}, \binits{S.}}:
\bctitle{A diffusion model framework for unsupervised neural combinatorial
  optimization}.
In: \bbtitle{International Conference on Machine Learning (ICML)}
(\byear{2024})
\end{bchapter}
\endbibitem

\bibitem[\protect\citeauthoryear{Lucas}{2014}]{lucas2014ising}
\begin{barticle}
\bauthor{\bsnm{Lucas}, \binits{A.}}:
\batitle{Ising formulations of many np problems}.
\bjtitle{Frontiers in Physics}
\bvolume{2},
\bfpage{5}
(\byear{2014})
\doiurl{10.3389/fphy.2014.00005}
\end{barticle}
\endbibitem

\bibitem[\protect\citeauthoryear{Farhi et~al.}{2014}]{farhi2014qaoa}
\begin{botherref}
\oauthor{\bsnm{Farhi}, \binits{E.}},
\oauthor{\bsnm{Goldstone}, \binits{J.}},
\oauthor{\bsnm{Gutmann}, \binits{S.}}:
A Quantum Approximate Optimization Algorithm
(2014)
\end{botherref}
\endbibitem

\bibitem[\protect\citeauthoryear{Ebadi et~al.}{2022}]{ebadi2022mis}
\begin{barticle}
\bauthor{\bsnm{Ebadi}, \binits{S.}},
\bauthor{\bsnm{Keesling}, \binits{A.}},
\bauthor{\bsnm{Cain}, \binits{M.}},
\bauthor{\bsnm{Wang}, \binits{T.T.}},
\bauthor{\bsnm{Levine}, \binits{H.}},
\bauthor{\bsnm{Bluvstein}, \binits{D.}},
\bauthor{\bsnm{Semeghini}, \binits{G.}},
\bauthor{\bsnm{Omran}, \binits{A.}},
\bauthor{\bsnm{Liu}, \binits{J.}},
\bauthor{\bsnm{Samajdar}, \binits{R.}},
\bauthor{\bsnm{Luo}, \binits{X.-Z.}},
\bauthor{\bsnm{Nash}, \binits{B.}},
\bauthor{\bsnm{Gao}, \binits{X.}},
\bauthor{\bsnm{Barak}, \binits{B.}},
\bauthor{\bsnm{Farhi}, \binits{E.}},
\bauthor{\bsnm{Sachdev}, \binits{S.}},
\bauthor{\bsnm{Gemelke}, \binits{N.}},
\bauthor{\bsnm{Zhou}, \binits{L.}},
\bauthor{\bsnm{Choi}, \binits{S.}},
\bauthor{\bsnm{Pichler}, \binits{H.}},
\bauthor{\bsnm{Wang}, \binits{S.}},
\bauthor{\bsnm{Greiner}, \binits{M.}},
\bauthor{\bsnm{Vuletic}, \binits{V.}},
\bauthor{\bsnm{Lukin}, \binits{M.D.}}:
\batitle{Quantum optimization of maximum independent set using rydberg atom
  arrays}.
\bjtitle{Science}
\bvolume{376}(\bissue{6598}),
\bfpage{1209}--\blpage{1215}
(\byear{2022})
\end{barticle}
\endbibitem

\bibitem[\protect\citeauthoryear{Cazals et~al.}{2025}]{cazals2025hard_mis}
\begin{botherref}
\oauthor{\bsnm{Cazals}, \binits{P.}},
\oauthor{\bsnm{Fran{\c{c}}ois}, \binits{A.}},
\oauthor{\bsnm{Henriet}, \binits{L.}},
\oauthor{\bsnm{Leclerc}, \binits{L.}},
\oauthor{\bsnm{Marin}, \binits{M.}},
\oauthor{\bsnm{Naghmouchi}, \binits{Y.}},
\oauthor{\bsnm{Silva~Coelho}, \binits{W.}},
\oauthor{\bsnm{Sikora}, \binits{F.}},
\oauthor{\bsnm{Vitale}, \binits{V.}},
\oauthor{\bsnm{Watrigant}, \binits{R.}},
\oauthor{\bsnm{Garzillo}, \binits{M.W.}},
\oauthor{\bsnm{Dalyac}, \binits{C.}}:
Identifying hard native instances for the maximum independent set problem on
  neutral atoms quantum processors
(2025)
\end{botherref}
\endbibitem

\bibitem[\protect\citeauthoryear{Wybo et~al.}{2026}]{wybo2026quantum_greedy}
\begin{botherref}
\oauthor{\bsnm{Wybo}, \binits{E.}},
\oauthor{\bsnm{R{\"o}nkk{\"o}}, \binits{J.}},
\oauthor{\bsnm{Hirviniemi}, \binits{O.}},
\oauthor{\bsnm{Fin{\v{z}}gar}, \binits{J.R.}},
\oauthor{\bsnm{Leib}, \binits{M.}}:
A scalable quantum-enhanced greedy algorithm for maximum independent set
  problems
(2026)
\end{botherref}
\endbibitem

\bibitem[\protect\citeauthoryear{Kumar et~al.}{2025}]{kumar2025trapped_ion_mis}
\begin{barticle}
\bauthor{\bsnm{Kumar}, \binits{S.}}, \betal:
\batitle{Digital-analog counterdiabatic quantum optimization with trapped
  ions}.
\bjtitle{Quantum Science and Technology}
\bvolume{10},
\bfpage{015023}
(\byear{2025})
\end{barticle}
\endbibitem

\bibitem[\protect\citeauthoryear{Nguyen et~al.}{2024}]{nguyen2024entropy}
\begin{botherref}
\oauthor{\bsnm{Nguyen}, \binits{L.}},
\oauthor{\bsnm{Miri}, \binits{M.-A.}},
\oauthor{\bsnm{Rupert}, \binits{R.J.}},
\oauthor{\bsnm{Dyk}, \binits{W.}},
\oauthor{\bsnm{Wu}, \binits{S.}},
\oauthor{\bsnm{Vrahoretis}, \binits{N.}},
\oauthor{\bsnm{Huang}, \binits{I.}},
\oauthor{\bsnm{Begliarbekov}, \binits{M.}},
\oauthor{\bsnm{Chancellor}, \binits{N.}},
\oauthor{\bsnm{Chukwu}, \binits{U.}},
\oauthor{\bsnm{Mahamuni}, \binits{P.}},
\oauthor{\bsnm{Martinez-Delgado}, \binits{C.}},
\oauthor{\bsnm{Haycraft}, \binits{D.}},
\oauthor{\bsnm{Spear}, \binits{C.}},
\oauthor{\bsnm{Campanelli}, \binits{M.}},
\oauthor{\bsnm{Huffman}, \binits{R.}},
\oauthor{\bsnm{Sua}, \binits{Y.M.}},
\oauthor{\bsnm{Huang}, \binits{Y.}}:
Entropy Computing: A Paradigm for Optimization in an Open Quantum System
(2024)
\end{botherref}
\endbibitem

\bibitem[\protect\citeauthoryear{Verstraete
  et~al.}{2009}]{verstraete2009dissipation}
\begin{barticle}
\bauthor{\bsnm{Verstraete}, \binits{F.}},
\bauthor{\bsnm{Wolf}, \binits{M.M.}},
\bauthor{\bsnm{Cirac}, \binits{J.I.}}:
\batitle{Quantum computation and quantum-state engineering driven by
  dissipation}.
\bjtitle{Nature Physics}
\bvolume{5}(\bissue{9}),
\bfpage{633}--\blpage{636}
(\byear{2009})
\end{barticle}
\endbibitem

\bibitem[\protect\citeauthoryear{Kraus et~al.}{2008}]{kraus2008quantum_markov}
\begin{barticle}
\bauthor{\bsnm{Kraus}, \binits{B.}},
\bauthor{\bsnm{B{\"u}chler}, \binits{H.P.}},
\bauthor{\bsnm{Diehl}, \binits{S.}},
\bauthor{\bsnm{Kantian}, \binits{A.}},
\bauthor{\bsnm{Micheli}, \binits{A.}},
\bauthor{\bsnm{Zoller}, \binits{P.}}:
\batitle{Preparation of entangled states by quantum markov processes}.
\bjtitle{Physical Review A}
\bvolume{78},
\bfpage{042307}
(\byear{2008})
\end{barticle}
\endbibitem

\bibitem[\protect\citeauthoryear{Wiseman and
  Milburn}{2010}]{wiseman2010quantum}
\begin{bbook}
\bauthor{\bsnm{Wiseman}, \binits{H.M.}},
\bauthor{\bsnm{Milburn}, \binits{G.J.}}:
\bbtitle{Quantum Measurement and Control}.
\bpublisher{Cambridge University Press},
\blocation{Cambridge}
(\byear{2010})
\end{bbook}
\endbibitem

\bibitem[\protect\citeauthoryear{Hungerford and
  Rinaldi}{2017}]{hungerford2017generalregularizedcontinuousformulation}
\begin{botherref}
\oauthor{\bsnm{Hungerford}, \binits{J.T.}},
\oauthor{\bsnm{Rinaldi}, \binits{F.}}:
A General Regularized Continuous Formulation for the Maximum Clique Problem
(2017)
\end{botherref}
\endbibitem

\bibitem[\protect\citeauthoryear{Johnson and Trick}{1996}]{johnson1996cliques}
\begin{bbook}
\beditor{\bsnm{Johnson}, \binits{D.S.}},
\beditor{\bsnm{Trick}, \binits{M.A.}} (eds.):
\bbtitle{Cliques, Coloring, and Satisfiability: Second DIMACS Implementation
  Challenge}.
\bsertitle{DIMACS Series in Discrete Mathematics and Theoretical Computer
  Science},
vol. \bseriesno{26}.
\bpublisher{American Mathematical Society},
\blocation{Providence, RI}
(\byear{1996})
\end{bbook}
\endbibitem

\bibitem[\protect\citeauthoryear{Armijo}{1966}]{Armijo1966MinimizationOF}
\begin{barticle}
\bauthor{\bsnm{Armijo}, \binits{L.}}:
\batitle{Minimization of functions having lipschitz continuous first partial
  derivatives.}
\bjtitle{Pacific Journal of Mathematics}
\bvolume{16},
\bfpage{1}--\blpage{3}
(\byear{1966})
\end{barticle}
\endbibitem

\bibitem[\protect\citeauthoryear{Birgin et~al.}{2000}]{birgin2000spg}
\begin{barticle}
\bauthor{\bsnm{Birgin}, \binits{E.G.}},
\bauthor{\bsnm{Mart{\'i}nez}, \binits{J.M.}},
\bauthor{\bsnm{Raydan}, \binits{M.}}:
\batitle{Nonmonotone spectral projected gradient methods on convex sets}.
\bjtitle{SIAM Journal on Optimization}
\bvolume{10}(\bissue{4}),
\bfpage{1196}--\blpage{1211}
(\byear{2000})
\end{barticle}
\endbibitem

\bibitem[\protect\citeauthoryear{{Hexaly}}{2026}]{Hexaly15}
\begin{botherref}
\oauthor{\bsnm{{Hexaly}}}:
Hexaly Optimizer.
\url{https://www.hexaly.com}
\end{botherref}
\endbibitem

\bibitem[\protect\citeauthoryear{Fortunato}{2010}]{fortunato2010community}
\begin{barticle}
\bauthor{\bsnm{Fortunato}, \binits{S.}}:
\batitle{Community detection in graphs}.
\bjtitle{Physics Reports}
\bvolume{486}(\bissue{3-5}),
\bfpage{75}--\blpage{174}
(\byear{2010})
{\href{https://arxiv.org/abs/0906.0612}{{arXiv:0906.0612}}}
{[physics.soc-ph]}
\end{barticle}
\endbibitem

\bibitem[\protect\citeauthoryear{Inagaki et~al.}{2016}]{inagaki2016cim}
\begin{barticle}
\bauthor{\bsnm{Inagaki}, \binits{T.}},
\bauthor{\bsnm{Haribara}, \binits{Y.}},
\bauthor{\bsnm{Igarashi}, \binits{K.}},
\bauthor{\bsnm{Sonobe}, \binits{T.}},
\bauthor{\bsnm{Tamate}, \binits{S.}},
\bauthor{\bsnm{Honjo}, \binits{T.}},
\bauthor{\bsnm{Marandi}, \binits{A.}},
\bauthor{\bsnm{McMahon}, \binits{P.L.}},
\bauthor{\bsnm{Umeki}, \binits{T.}},
\bauthor{\bsnm{Enbutsu}, \binits{K.}},
\bauthor{\bsnm{Tadanaga}, \binits{O.}},
\bauthor{\bsnm{Takenouchi}, \binits{H.}},
\bauthor{\bsnm{Aihara}, \binits{K.}},
\bauthor{\bsnm{Kawarabayashi}, \binits{K.-i.}},
\bauthor{\bsnm{Inoue}, \binits{K.}},
\bauthor{\bsnm{Utsunomiya}, \binits{S.}},
\bauthor{\bsnm{Takesue}, \binits{H.}}:
\batitle{A coherent ising machine for 2000-node optimization problems}.
\bjtitle{Science}
\bvolume{354}(\bissue{6312}),
\bfpage{603}--\blpage{606}
(\byear{2016})
\end{barticle}
\endbibitem

\bibitem[\protect\citeauthoryear{McMahon et~al.}{2016}]{mcmahon2016cim}
\begin{barticle}
\bauthor{\bsnm{McMahon}, \binits{P.L.}},
\bauthor{\bsnm{Marandi}, \binits{A.}},
\bauthor{\bsnm{Haribara}, \binits{Y.}},
\bauthor{\bsnm{Hamerly}, \binits{R.}},
\bauthor{\bsnm{Langrock}, \binits{C.}},
\bauthor{\bsnm{Tamate}, \binits{S.}},
\bauthor{\bsnm{Inagaki}, \binits{T.}},
\bauthor{\bsnm{Takesue}, \binits{H.}},
\bauthor{\bsnm{Utsunomiya}, \binits{S.}},
\bauthor{\bsnm{Aihara}, \binits{K.}},
\bauthor{\bsnm{Byer}, \binits{R.L.}},
\bauthor{\bsnm{Fejer}, \binits{M.M.}},
\bauthor{\bsnm{Mabuchi}, \binits{H.}},
\bauthor{\bsnm{Yamamoto}, \binits{Y.}}:
\batitle{A fully-programmable 100-spin coherent ising machine with all-to-all
  connections}.
\bjtitle{Science}
\bvolume{354}(\bissue{6312}),
\bfpage{615}--\blpage{617}
(\byear{2016})
\end{barticle}
\endbibitem

\bibitem[\protect\citeauthoryear{Yamamoto et~al.}{2020}]{yamamoto2020cim}
\begin{botherref}
\oauthor{\bsnm{Yamamoto}, \binits{Y.}},
\oauthor{\bsnm{Leleu}, \binits{T.}},
\oauthor{\bsnm{Ganguli}, \binits{S.}},
\oauthor{\bsnm{Mabuchi}, \binits{H.}}:
Coherent Ising machines -- Quantum optics and neural network perspectives
(2020)
\end{botherref}
\endbibitem

\bibitem[\protect\citeauthoryear{Yamamura
  et~al.}{2017}]{yamamura2017cim_quantum_model}
\begin{barticle}
\bauthor{\bsnm{Yamamura}, \binits{A.}},
\bauthor{\bsnm{Aihara}, \binits{K.}},
\bauthor{\bsnm{Yamamoto}, \binits{Y.}}:
\batitle{Quantum model for coherent ising machines: Discrete-time measurement
  feedback formulation}.
\bjtitle{Physical Review A}
\bvolume{96},
\bfpage{053834}
(\byear{2017})
\end{barticle}
\endbibitem

\bibitem[\protect\citeauthoryear{Takesue et~al.}{2025}]{takesue2025cim_mis}
\begin{barticle}
\bauthor{\bsnm{Takesue}, \binits{H.}},
\bauthor{\bsnm{Inaba}, \binits{K.}},
\bauthor{\bsnm{Honjo}, \binits{T.}},
\bauthor{\bsnm{Yamada}, \binits{Y.}},
\bauthor{\bsnm{Ikuta}, \binits{T.}},
\bauthor{\bsnm{Yonezu}, \binits{Y.}},
\bauthor{\bsnm{Inagaki}, \binits{T.}},
\bauthor{\bsnm{Umeki}, \binits{T.}},
\bauthor{\bsnm{Kasahara}, \binits{R.}}:
\batitle{Finding independent sets in large-scale graphs with a coherent {Ising}
  machine}.
\bjtitle{Science Advances}
\bvolume{11}(\bissue{7}),
\bfpage{7223}
(\byear{2025})
\end{barticle}
\endbibitem

\bibitem[\protect\citeauthoryear{Jagota and Pelillo}{1995}]{jagota1995feasible}
\begin{botherref}
\oauthor{\bsnm{Jagota}, \binits{A.}},
\oauthor{\bsnm{Pelillo}, \binits{M.}}:
Feasible and infeasible maxima in a quadratic program for maximum clique.
Journal of Artificial Neural Networks
\textbf{2}
(1995)
\end{botherref}
\endbibitem

\bibitem[\protect\citeauthoryear{Gibbons et~al.}{1996}]{gibbons1996continuous}
\begin{barticle}
\bauthor{\bsnm{Gibbons}, \binits{L.E.}},
\bauthor{\bsnm{Hearn}, \binits{D.W.}},
\bauthor{\bsnm{Pardalos}, \binits{P.M.}}:
\batitle{A continuous based heuristic for the maximum clique problem}.
\bjtitle{DIMACS Series in Discrete Mathematics and Theoretical Computer
  Science}
\bvolume{26},
\bfpage{103}--\blpage{124}
(\byear{1996})
\end{barticle}
\endbibitem

\bibitem[\protect\citeauthoryear{Grosso et~al.}{2008}]{grosso2008heuristics}
\begin{barticle}
\bauthor{\bsnm{Grosso}, \binits{A.}},
\bauthor{\bsnm{Locatelli}, \binits{M.}},
\bauthor{\bsnm{Pullan}, \binits{W.}}:
\batitle{Simple ingredients leading to very efficient heuristics for the
  maximum clique problem}.
\bjtitle{Journal of Heuristics}
\bvolume{14}(\bissue{6}),
\bfpage{587}--\blpage{612}
(\byear{2008})
\end{barticle}
\endbibitem

\bibitem[\protect\citeauthoryear{Wu and Hao}{2015}]{WU2015693}
\begin{barticle}
\bauthor{\bsnm{Wu}, \binits{Q.}},
\bauthor{\bsnm{Hao}, \binits{J.-K.}}:
\batitle{A review on algorithms for maximum clique problems}.
\bjtitle{European Journal of Operational Research}
\bvolume{242}(\bissue{3}),
\bfpage{693}--\blpage{709}
(\byear{2015})
\end{barticle}
\endbibitem

\bibitem[\protect\citeauthoryear{McCreesh and Prosser}{2013}]{a6040618}
\begin{barticle}
\bauthor{\bsnm{McCreesh}, \binits{C.}},
\bauthor{\bsnm{Prosser}, \binits{P.}}:
\batitle{Multi-threading a state-of-the-art maximum clique algorithm}.
\bjtitle{Algorithms}
\bvolume{6}(\bissue{4}),
\bfpage{618}--\blpage{635}
(\byear{2013})
\end{barticle}
\endbibitem

\bibitem[\protect\citeauthoryear{Karalias and
  Loukas}{2021}]{karalias2021erdos_neural}
\begin{botherref}
\oauthor{\bsnm{Karalias}, \binits{N.}},
\oauthor{\bsnm{Loukas}, \binits{A.}}:
Erdos Goes Neural: an Unsupervised Learning Framework for Combinatorial
  Optimization on Graphs
(2021)
\end{botherref}
\endbibitem

\bibitem[\protect\citeauthoryear{Acikalin et~al.}{2025}]{acikalin2025x2gnn}
\begin{bchapter}
\bauthor{\bsnm{Acikalin}, \binits{U.U.}},
\bauthor{\bsnm{Ferber}, \binits{A.M.}},
\bauthor{\bsnm{Gomes}, \binits{C.P.}}:
\bctitle{Learning to explore and exploit with {GNNs} for unsupervised
  combinatorial optimization}.
In: \bbtitle{International Conference on Learning Representations (ICLR)}
(\byear{2025})
\end{bchapter}
\endbibitem

\bibitem[\protect\citeauthoryear{Duchi et~al.}{2008}]{duchi2008efficient}
\begin{bchapter}
\bauthor{\bsnm{Duchi}, \binits{J.}},
\bauthor{\bsnm{Shalev-Shwartz}, \binits{S.}},
\bauthor{\bsnm{Singer}, \binits{Y.}},
\bauthor{\bsnm{Chandra}, \binits{T.}}:
\bctitle{Efficient projections onto the l1-ball for learning in high
  dimensions}.
In: \bbtitle{Proceedings of the 25th International Conference on Machine
  Learning},
pp. \bfpage{272}--\blpage{279}
(\byear{2008}).
\bcomment{ACM}
\end{bchapter}
\endbibitem

\bibitem[\protect\citeauthoryear{Bradbury et~al.}{2018}]{jax2018github}
\begin{botherref}
\oauthor{\bsnm{Bradbury}, \binits{J.}},
\oauthor{\bsnm{Frostig}, \binits{R.}},
\oauthor{\bsnm{Hawkins}, \binits{P.}},
\oauthor{\bsnm{Johnson}, \binits{M.J.}},
\oauthor{\bsnm{Leary}, \binits{C.}},
\oauthor{\bsnm{Maclaurin}, \binits{D.}},
\oauthor{\bsnm{Necula}, \binits{G.}},
\oauthor{\bsnm{Paszke}, \binits{A.}},
\oauthor{\bsnm{VanderPlas}, \binits{J.}},
\oauthor{\bsnm{Wanderman-Milne}, \binits{S.}},
\oauthor{\bsnm{Zhang}, \binits{Q.}}:
{JAX}: composable transformations of {P}ython+{N}um{P}y programs
(2018)
\end{botherref}
\endbibitem

\bibitem[\protect\citeauthoryear{Pelillo}{1999}]{pelillo1999replicator}
\begin{barticle}
\bauthor{\bsnm{Pelillo}, \binits{M.}}:
\batitle{Replicator equations, maximal cliques, and graph isomorphism}.
\bjtitle{Neural Computation}
\bvolume{11}(\bissue{8}),
\bfpage{1933}--\blpage{1955}
(\byear{1999})
\end{barticle}
\endbibitem

\bibitem[\protect\citeauthoryear{Bomze et~al.}{2002}]{bomze2002annealed}
\begin{barticle}
\bauthor{\bsnm{Bomze}, \binits{I.M.}},
\bauthor{\bsnm{Budinich}, \binits{M.}},
\bauthor{\bsnm{Pelillo}, \binits{M.}},
\bauthor{\bsnm{Rossi}, \binits{C.}}:
\batitle{Annealed replication: a new heuristic for the maximum clique problem}.
\bjtitle{Discrete Applied Mathematics}
\bvolume{121},
\bfpage{27}--\blpage{49}
(\byear{2002})
\end{barticle}
\endbibitem

\bibitem[\protect\citeauthoryear{Kadowaki and
  Nishimori}{1998}]{kadowaki1998quantum_annealing}
\begin{barticle}
\bauthor{\bsnm{Kadowaki}, \binits{T.}},
\bauthor{\bsnm{Nishimori}, \binits{H.}}:
\batitle{Quantum annealing in the transverse ising model}.
\bjtitle{Physical Review E}
\bvolume{58},
\bfpage{5355}
(\byear{1998})
\end{barticle}
\endbibitem

\bibitem[\protect\citeauthoryear{Oh et~al.}{2024}]{oh2024gbs_classical}
\begin{barticle}
\bauthor{\bsnm{Oh}, \binits{C.}},
\bauthor{\bsnm{Fefferman}, \binits{B.}},
\bauthor{\bsnm{Jiang}, \binits{L.}},
\bauthor{\bsnm{Quesada}, \binits{N.}}:
\batitle{Quantum-inspired classical algorithm for graph problems by {Gaussian}
  boson sampling}.
\bjtitle{PRX Quantum}
\bvolume{5},
\bfpage{020341}
(\byear{2024})
\end{barticle}
\endbibitem

\bibitem[\protect\citeauthoryear{Gherardi and
  Leporati}{2024}]{gherardi2024dwave_maxclique}
\begin{botherref}
\oauthor{\bsnm{Gherardi}, \binits{A.}},
\oauthor{\bsnm{Leporati}, \binits{A.}}:
An Analysis of Quantum Annealing Algorithms for Solving the Maximum Clique
  Problem
(2024)
\end{botherref}
\endbibitem

\bibitem[\protect\citeauthoryear{Schuetz et~al.}{2025}]{schuetz2025qredumis}
\begin{botherref}
\oauthor{\bsnm{Schuetz}, \binits{M.J.A.}}, et al.:
{qReduMIS}: A Quantum-Informed Reduction Algorithm for the Maximum Independent
  Set Problem
(2025)
\end{botherref}
\endbibitem

\end{thebibliography}

\end{document}